# Lithium Diffusion in Li$_2$X(X=O, S and Se): Ab-initio Simulations and Neutron Inelastic Scattering Measurements

M. K. Gupta[1*], Baltej Singh[1,2], Prabhatasree Goel[1], R. Mittal[1,2#], S. Rols[3] and S. L. Chaplot[1,2]

[1]Solid State Physics Division, Bhabha Atomic Research Centre, Mumbai, 400085, India
[2]Homi Bhabha National Institute, Anushaktinagar, Mumbai 400094, India
[3]Institut Laue-Langevin, 71 avenue des Martyrs, Grenoble Cedex 9, 38042, France
Corresponding Authors Email: mayankg@barc.gov.in*, rmittal@barc.gov.in#

We have performed ab-initio lattice dynamics and molecular dynamics studies of Li$_2$X (X=O, S and Se) to understand the ionic conduction in these compounds. The inelastic neutron scattering measurements on Li$_2$O have been performed across its superionic transition temperature of about 1200 K. The experimental spectra show significant changes around the superionic transition temperature, which is attributed to large diffusion of lithium as well as its large vibrational amplitude. We have identified a correlation between the chemical pressure (ionic radius of X atom) and the superionic transition temperature. The simulations are able to provide the ionic diffusion pathways in Li$_2$X.

# I. Introduction

Solid state batteries are highly sustainable and stable[1, 2]. The use of solid electrolyte between the electrodes has enabled to solve the issue of dendrite formation and leakage of electrolyte [1, 3]. An extensive amount of research is underway to predict and design solid ionic conductors with ionic conductivity in the range of that of liquids[4-19]. They are called the superionic conductors and possess an extremely low potential energy barrier for diffusion in the solid-state. Li ion batteries play a very important role in hybrid electric vehicle and large grid technology[20, 21]. Ionic conductivity of electrode and electrolyte materials is an important parameter for the operation of Li-based batteries[20, 22]. The diffusion coefficient of $Li_2S$ and $Li_2O$ has been found to be $1.39 \times 10^{-9}$ $m^2$/sec [23] and $1.19 \times 10^{-9}$ $m^2$/sec [24] at elevated temperature respectively. Such compounds find application in photocathodes, fuel cells, power sources, UV space technology appliances etc.

There is a fundamental difference between the conventional ionic conductors and superionic conductors. Superionic conductors show sudden jump in ionic conductivity above certain temperature in contrast to conventional ionic conductor [4-19]. Further, in case of superionic conductors, there is a clear indication of a phase change either sharp or diffuse, during which conductivity rises[15, 25, 26]. Li ion diffuses while the host lattice (oxygen/sulphur/selnium) ions constitute the rigid framework. The mechanism and rate of Li diffusion through Li-based battery materials have been studied extensively [21, 27, 28]. In addition, different strategies have been proposed to increase the carrier concentration and mobility in order to improve the Li ion conductivity in solid state materials [28, 29]. The search of superionic conductor $Li_{10}GeP_2S_{12}$ (LGPS) by Kamaya, *et al*[14], triggered the attention of material scientists to activate the field of Li and Na superionic conductors. Many oxides and sulphide compounds like $Li_7La_3Zr_2O_{12}$[6] and NASICON[30, 31] discovered later, were found to exhibit high Li ionic conductivity of ∼1–10 mS $cm^{-1}$ at room temperature (RT), and low activation energy, ∼0.2–0.3 eV.

Above the superionic transition temperature, when diffusion of a particular species of ions occurs inside a solid, the sub-lattice of those ions melts, which may be accompanied by softening of the related phonon modes[32-34]. The eigenvector of these soft modes can be related to the directions of ionic diffusion and hence might suggest the minimum energy pathways for diffusion[34, 35]. The static energy transition calculations along those specific pathways can be used for estimation of the activation energy barrier. Understanding the role of anharmonic phonon modes responsible for the diffusion mechanism is very important for improving ionic conduction.

Extensive experimental and theoretical studies have been carried out to understand the structure and diffusion behaviour of superionic compounds[17, 36, 37]. X-ray diffraction, neutron scattering, dielectric and NMR spectroscopic techniques proved useful for characterizing the structure and ionic conduction in the battery materials[38-41]. However, the underlying mechanism of atomic level behavior of atoms giving rise to these properties can be studied well by ab initio computational techniques[29, 42]. Recent advancement in the computer hardware has enabled material scientists to perform ab-initio molecular dynamics on hundreds of atoms. The calculations have been successfully applied to many superionic conductors to predict and understand anisotropic diffusion pathways and correlated ionic movements[3, 15, 29, 43, 44]. Quasielastic neutron scattering studies on diffusion in $Li_2S$ by Altorfer et al [23] discusses two models for Li hopping in the fast ion conduction phase, namely (i) Li jumping between vacancy and a regular site; (ii) Li jumping between interstitial and regular sites.

Li-based compounds are of special interest due to their applications in Li-based batteries. Simple fluorite/antifluorite structured compounds mostly exhibit superionic transition. At ambient conditions $Li_2S$ and $Li_2O$ occur in the anti-fluorite structure with space group $O_h^5$ (Fm3m) [45, 46]. Sulphur and oxygen ions are arranged in a FCC sublattice with lithium ions occupying the tetrahedral sites. The compounds undergo pressure induced transformation from anti-fluorite to anti-cottunite phase at high pressure of 13 GPa ($Li_2S$) and 45 GPa ($Li_2O$) respectively [47, 48]. Copious simulation studies have been carried out on similar fluorites and antifluorites like $CaF_2$, $UO_2$, $Li_2O$ etc. We have previously carried out extensive studies on the superionic property of $Li_2O$ using theoretical simulations to deduce the possible easy direction for the diffusion of lithium [33]. Several glassy solutions of $Li_2S$ and $Li_2O$ have been studied extensively in $Li_2S-P_2S_5$ [49], and $Li_2O-B_2O_3-SiO_2$ with search for increased ionic conduction in mind[50]. There are indirect Raman[51] studies which indicate that $Li_2S$ shows fast ion conduction at 850 K. There are no reported data on measurement of diffusion coefficient in the compounds. A thorough investigation of temperature-dependent Li ion conductivity studies in $Li_2S$ is clearly needed.

Here we focus our study to investigate the factors which trigger superionic behaviour in these Li-based fluorite structures. We have chosen $Li_2O$, $Li_2S$ and $Li_2Se$ compounds. These are fast ion conductors exhibiting high ionic conductivity at high temperatures well below the melting temperature. Due to large electronic band gap they may find applications in battery material as electrolyte. But the thermodynamic and mechanical stability of electrolyte is another factor to be taken care. The mechanical and thermodynamical stabilities are governed by the phonons in crystalline materials. Here, in this work our aim is to understand the role of lattice dynamics in diffusion of Li ion and the stability of crystal in these fluorite structures. Our main impetus in this work is to carry out a comprehensive study of vibrational properties of the compounds along with high temperature molecular dynamics simulations to understand the superionic conduction mechanism in $Li_2O$, $Li_2S$ and $Li_2Se$. Our experimental neutron scattering studies reveals the behavior of Li sub-lattice of $Li_2O$ with increasing temperatures. This would give us an idea into the possible role of phonons in the high temperature behavior of these compounds. In this manner, we would be able to throw some light on the mechanism of diffusion in $Li_2O$, $Li_2S$ and $Li_2Se$.

**II. Experimental**

The polycrystalline samples of $Li_2O$ were purchased from Sigma Aldrich. The inelastic neutron scattering experiment on $Li_2O$ is carried out using the IN4C spectrometer at the Institut Laue Langevin (ILL), France. The spectrometer is based on the time-of-flight technique and is equipped with a large detector bank covering a wide range of about $10^o$ to $110^o$ of scattering angle. The inelastic neutron scattering measurements were performed at several temperatures from 300 K to 1273 K. About 2 cc of polycrystalline sample of $Li_2O$ has been used for the measurements. For these measurements we have used an incident neutron wavelength of 2.4 Å (14.2 meV) in neutron energy gain setup. In the incoherent one-phonon approximation, the measured scattering function $S(Q,E)$, as observed in the neutron experiments, is related[52-54] to the phonon density of states $g^{(n)}(E)$ as follows:

$$g^{(n)}(E) = A < \frac{e^{2W(Q)}}{Q^2} \frac{E}{n(E,T) + \frac{1}{2} \pm \frac{1}{2}} S(Q,E) > \qquad (1)$$

$$g^n(E) = B \sum_k \{\frac{4\pi b_k^2}{m_k}\} g_k(E) \qquad (2)$$

where the + or − signs correspond to energy loss or gain of the neutrons respectively and where $n(E,T) = [\exp(E/k_BT) - 1]^{-1}$. $A$ and $B$ are normalization constants and $b_k$, $m_k$, and $g_k(E)$ are, respectively, the neutron scattering length, mass, and partial density of states of the $k^{th}$ atom in the unit cell. The quantity between $<>$ represents suitable average over all $Q$ values at a given energy. $2W(Q)$ is the Debye-Waller factor averaged over all the atoms. The weighting factors $\frac{4\pi b_k^2}{m_k}$ for various atoms in the units of barns/amu are: 0.1974 and 0.2645 for Li and O respectively. The values of neutron scattering lengths for various atoms can be found from Ref.[55].

## III. Computational Details

The lattice and molecular dynamics simulations of $Li_2X$ (X=O, S and Se) are performed in the cubic phase (space group Fm-3m) using ab-initio density functional theory as implemented in the VASP software [56, 57]. A supercell of (2 × 2 × 2) dimension, which consist of 96 atoms has been used in the computations. In the lattice dynamics calculations, the required force constants were computed within the Hellman-Feynman framework, on various atoms in different configurations of a supercell with (±x, ±y, ±z) atomic displacement patterns. An energy cut-off of 900 eV was used for plane wave expansion. The Monkhorst Pack method[58] is used for k point generation with a 2×2×2 k-point mesh. The valence electronic configurations of Li and X (O, S and Se), as used in calculations for pseudo-potential generation are $1s^2 2s^1$ and $s^2 p^4$ respectively. The convergence breakdown criteria for the total energy and ionic force loops were set to $10^{-8}$ eV and $10^{-4}$ eV Å$^{-1}$, respectively. We have used phonopy software[59] to obtain the phonon frequencies in the entire Brillouin zone, as a subsequent step to density functional theory total energy calculations.

The thermal expansion behaviour has been computed under the quasiharmonic approximation[60]. Each phonon mode of energy $E_{qj}$ ($j^{th}$ phonon mode at point q in the Brillouin zone) contributes to the thermal expansion coefficient, which is given by the following relation :

$$\alpha_V(T) = \frac{1}{BV_0} \sum_{q,j} C_v(q,j,T) \Gamma_{q,j} \qquad (3)$$

Where $V_0$ is the unit cell volume, $\Gamma_{q,j}$ is the mode Grüneisen parameter, $C_v(q,j,T)$ is the specific-heat contribution of the phonons of energy $E_{q,j}$. The mode Grüneisen parameter of the phonon of energy $E_{q,j}$ is given as,

$$\Gamma_{q,j} = -\left(\frac{\partial lnE_{q,j}}{\partial lnV}\right) \qquad (4)$$

The molecular dynamics simulations are carried out in NVE ensemble for 40 pico-second, with a time step of 1femtosecond. The temperature in the NVT simulations is attained through a Nose thermostat[61]. Initially, the structure was equilibrated for 10 ps to attain the required temperature in NVT simulations. Then the production runs up to 40 ps, NVE simulations are performed. Simulations are performed for a series of temperatures from 300 to 1600 K. At each temperature, a well-equilibrated configuration is observed during the 40 ps simulation.

## IV. Results and Discussion

### A. Temperature Dependent Measurement of Phonon Spectra

We have performed the inelastic neutron scattering measurements of Li$_2$O at several temperatures (**Fig 1(a)**) from 300 K to 1273K covering the superionic transition at about 1200 K. The room temperature measurements show well defined peak structure in the phonon density of states and it disappears at above 973K below the superionic transition temperature. We have compared the measured phonon density of states at 300 K with the lattice dynamics calculated phonon spectrum (**Fig 1(b)**). The neutron-weighted phonon density of states show peaks at about 30 meV, 50meV , 65 meV and 90 meV. Partial contributions from Li and O atoms to the total density of states are also shown. We could see that both Li and O have almost equal contribution in entire spectrum except at about 50-70 meV regime, where Li have dominating contribution. The calculated total neutron weighted phonon density of states is in good agreement with measurements (**Fig 1(b)**). As temperature is increased to 773 K, the peak at about 50 meV in the phonon density of states is found to shift toward lower energy with respect to the spectrum at 300 K(Fig 1(a)). This could well be understood by comparing the partial contribution of lithium and oxygen with the measured phonon density of states. At higher temperature the mean squared amplitude of Li vibration becomes very large, and their contribution to the total density of states ( $g_k(E)e^{-Q^2u^2}$ ) becomes smaller and contribution from oxygen will dominate. Further at higher temperature, phonon spectrum of oxygen sub-lattice softens due to large thermal expansion of Li$_2$O. On further increase of temperature to 973 K and above, the measured spectra seem like a broad envelope, this could be due to large mean square displacement of Li and O as well as diffusion of Li.

At high temperature, where the anharmonicity of vibrations becomes significant, we need to perform molecular dynamics simulation to evaluate the phonon spectrum. In **Fig 2**, we have shown the calculated phonon density of states from molecular dynamics simulations at 300 K and 1200 K and compared it with the measurements. The room temperature calculated phonon density of states reproduces the peak features close to the measurements, whereas at 1200 K the spectrum gets significantly broadened. The fair agreement between the calculation and measurements validate the robustness of the method and can be utilized for further microscopic analysis.

## B. Thermodynamics Behaviour of $Li_2X$ (X=O, S and Se) from Lattice Dynamics

In **Fig 3**, we have computed the phonon dispersion relation of $Li_2X$ (X=O, S and Se) along various high symmetry directions and compared with the available measurements[62, 63]. The phonon dispersion relation in $Li_2O$ shows that along (001) direction one of the optic branches shows large dispersion and at the zone boundary point the energy of this optic phonon branch is lowest (24 meV). Interestingly the eigenvector of this phonon mode shows displacement of lithium atoms while oxygen's remain at rest which is linked to phonon instability in $Li_2O$ at high temperatures[33]. However in $Li_2S$ and $Li_2Se$, the same optic branch shows less dispersion behaviour and also the phonon mode energy at zone boundary is higher (30 meV) than that in $Li_2O$ (24 meV). This may be change due to softer host structure of Se and S as compared with O host structure as well as the difference in interaction strength of Li with host lattice. It seems that as we increase the size of the cation X (X=O, S and Se) the dispersion of the optic branch become less dispersive.

In our previous work, we have shown[33] that with change in volume the optic mode at zone boundary becomes unstable when the volume increases to the value close to superionic transition temperature, indicating that volume driven instability of phonon mode may initiate the lithium diffusion in $Li_2O$ at higher temperature. We have calculated the volume dependence of zone boundary phonon mode at (001) in other two compounds (**Fig 4**). We can see that the percentage lattice parameter 'a' change required for phonon instability in $Li_2O$ is about 6.6%; however, in $Li_2S$ and $Li_2Se$ this is as large as 12.5% and 12.9% respectively, which is not possible. Such expansion of the lattice would result in melting of the compound. As mentioned above, it seems that in case of $Li_2O$ the lowest optical branch shows the large dispersion and the zone boundary mode of this branch destabilizes at higher temperature. While in other two compounds the same branch does not show significant dispersion and even at high temperature the zone boundary mode of the same may not destabilize. Further, due to lower Li band centre and soft host structure, a significant number of lithium are vibrating with higher amplitude and may lead to diffusion in $Li_2S$ and $Li_2Se$ at high temperature.

In **Fig 5**, we have shown the calculated partial and total density of states of $Li_2X$ (X=O, S and Se) compounds. The Li and O partial density of states in $Li_2O$ extend in the entire spectrum range up to 100 meV. However, in $Li_2S$ the spectral range extends up to 50 meV only. The lower part of the spectrum up to 30 meV is dominated by S, while higher energy spectrum above 30 meV has contributions mainly from Li. In $Li_2Se$, the contribution due to Li and Se are well separated. The contributions due to Se and Li are below and above 20 meV respectively. The band centre of the peaks of Li phonon density of states in various compounds are at 57 meV, 42 meV and 38 meV in $Li_2O$, $Li_2S$ and $Li_2Se$ respectively; and for O, S and Se the vibrations are centered at about 45 meV, 21meV and 12 meV respectively.

It can be seen that the Li band centre moves towards lower energy regime as we go from $Li_2O$ to $Li_2Se$, which indicates decrease in bonding strength of Li atom with host, which is good for better ionic conductivity. While the shift of the band centre of anion (X=O, S and Se) indicates the poor stability of host structure, i.e. the mechanical stability of O sub-lattice is good while that of Se sub-lattice is least among three compounds. This suggest that from stability point of view $Li_2O$ is best among the three, however, from diffusion point of view $Li_2Se$ is better than other two. So there is a trade-off between host stability and ionic conduction. This suggests that for a better electrolyte/electrode one can play with stoichiometry of O, S and Se to optimize the property in a mixture of all these compounds. By doing this, one may achieve the low energy Li spectrum and moderate host stability.

The fluorite structure compounds are also known for their high thermal expansion coefficient. The volume thermal expansion coefficient of $Li_2X$ compounds has been computed under quasiharmonic approximation (**Fig 6**). The calculated coefficient of volume thermal expansion at 400 K is $\sim 80 \times 10^{-6}$ $K^{-1}$ in all three compounds. However, at lower temperature the $Li_2Se$ shows the largest thermal expansion among all $Li_2X$ (X=O,S and Se). The lower band centre of host element X(O,S and Se) in $Li_2Se$ and $Li_2S$ in comparison to $Li_2O$, facilitated the host structure to expand at much lower temperature. The comparison between the experimental[64, 65] and calculated thermal expansion behavior of $Li_2O$ and $Li_2S$ is shown in Fig. 7. The deviation between experiments and calculations increases above 600 K, which indicates that at high temperature explicit anharmonic effects[60] due to large thermal amplitude of atoms are important. The quasiharmonic calculations of thermal expansion only consider the change in phonon energies with volume.

## C. Estimate of Barrier Energy for Lithium Migration

In order to get an estimate of barrier energy for lithium migration from one tetrahedral site to another tetrahedral site we have performed the energy barrier calculations along various high symmetry directions using nudged elastic band (NEB) methods. These calculations are performed for Li movements along [100], [110] and [111] directions. A number of structural images are generated along the initial path of diffusion which is then optimized to obtain the actual diffusion pathways and activation energy barrier. The calculated activation energy barrier and final optimized pathways are shown in **Fig 8.** It can be seen that the diffusion pathway along [100] direction in all $Li_2X$ compounds is nearly straight. We found that activation energy is least for $Li_2O$ compound along [100]. This implies that mechanism for Li diffusion along [100] is more probable in $Li_2O$ in comparison to that for $Li_2S$ and $Li_2Se$.

The second path for Li diffusion along [110] direction shows (**Fig 8**) curved diffusion pathway. The activation energy barriers for Li diffusion along this direction shows higher barrier for $Li_2O$ than that for $Li_2S$ and $Li_2Se$. This may arise from the more open channel along [110] in $Li_2Se$ and $Li_2S$ for Li diffusion than that in $Li_2O$ due to larger size of Se and S than O. The diffusion of Li along [111] seems (**Fig 8**) to be prohibited for $Li_2O$ due to large barrier energy. However in case of $Li_2S$ and $Li_2Se$, there is a significant attractive potential in a smaller vicinity towards <111> direction and thereafter it shows large barrier. Such potentials might lead to a possibility of complex diffusion pathways. Hence to examine other possibilities of Li migration, we have performed the analysis of Li trajectory obtained from molecular dynamics simulation (Section IVD).

Our molecular dynamics simulations shows very different and interesting pathways for diffusion in these compounds. The interesting thing to note in these calculations (**Fig 8**) is the dip along [111] directions at (¼ ¼ ¼) position of Li for $Li_2S$ and $Li_2Se$. This may be arising from the possibility of available jump sites at these intervening positions. The possibility of these intervening jumps increases as we go from $Li_2O$ to $Li_2Se$.

**D. Thermodynamics Properties of $Li_2X$ (X=O, S and Se) from Molecular Dynamics**

The above discussion is based on the lattice dynamics calculations which are performed at 0 K and the thermodynamic behaviour is obtained under the quasiharmonic approximation. The lattice dynamics provide us the estimate of the phonon properties (density of states, elastic properties etc) of materials and help us to understand the difference in material properties at 0 K. However, at higher temperature when system is close to some instability or transition (like superionic transition in this case), the effect of anharmonicity significantly dominates and cannot be accurately represented by quasiharmonic effects. Hence in order to account the anharmonic effect and understand the Li diffusion and its consequences on inelastic neutron scattering spectrum we have carried out the ab-intio molecular dynamics simulation. The calculations are performed at various temperatures ranging from room temperature to across the superionic transition temperature of $Li_2X$ compounds.

We have first looked at the effect of temperature on the vibrational spectrum i.e. phonon density of states of various elements in $Li_2X$ (**Fig 9**). We found that at room temperature the spectrum of Li and X (O, S and Se) shows sharp peak features. However, at 1200 K (above the superionic temperature) the Li spectrum becomes very broad and at zero energy there is finite density of states, which is a signature of diffusion. The magnitude of this shift is maximum in $Li_2Se$, which is again consistent with our lattice dynamics result that Li ion diffusion is much easier in $Li_2Se$ in comparison to the other two compounds. Further, we could see that the density of states of host element, i.e. X(X=O, S and Se) also broaden with temperature but lesser in magnitude than that of Li. The peak like structure in the spectra still persists above the superionic transition temperature, indicating that the atoms remain intact with the lattice most of the time.

Further, diffusion of Li occurs in the lattice at high temperature. Its local environment keeps changing with time, and this will give rise to broadening in the pair distribution function (PDF) of Li-X (X=O, S and Se) and Li-Li averaged over space and time. We have calculated (**Fig 10**) the pair distribution function for various pair of atoms in $Li_2X$ (X=O, S and Se). We could see that at room temperature, the PDF of all pairs of atoms show sharp peaks at finite positions. This is due to the fact that at room temperature the local environments of all the atoms are almost invariant with time. We find that the PDF of Li-Li and Li-X pairs at higher temperature of 1200 K clearly show only first two peaks (i.e. up to second neighbor distance) and the peaks at higher interatomic distance get broadened. The PDF between X-X (X=O, S and Se) atoms shows peak like structure up to a few neighbors. Hence the Li sub-lattice in $Li_2X$ melts at superionic temperature, while the host element X(X=O, S and Se) remain intact with the lattice. Similarly at higher temperature the bond angle in the $Li_4X$ polyhedral unit changes with time and it leads to broadening in the angle distribution (**Fig 11**).

**E.  Diffusion in $Li_2X$ (X=O, S and Se) from Molecular Dynamics**

In order to design a material with large diffusion coefficient, e.g., electrolyte in battery, one needs to know various channels for diffusion, easy directions of diffusion, activation energy barriers and stability of host structure etc. The mean square displacement is a very important quantity in this regard as it contains most of the information. Hence in order to understand and explain the difference in Li diffusion between various $Li_2X$ (X=O, S and Se) compounds, we have computed the mean square displacement of various atoms as a function of time at different temperatures (**Fig 12**). The mean square displacements are averaged over all the atoms of each type. At room temperature the magnitude of mean square displacement of various atoms in $Li_2X$ (X=O, S and Se) is independent with time. Since, the diffusion coefficient is proportional to the time derivative of mean square displacement (MSD), at room temperature the derivative of MSD with time for different element in $Li_2X$ is zero. Hence there is no diffusion at room temperature. As we go to high temperature close to the superionic regime, the mean square displacement of Li atoms in all three compounds starts increasing with time.

At superionic regime, the host structure is stable and provides discrete sites for Li occupancy, and Li can diffuse through these sites via jump from one site to another; hence the mean square displacement of individual atoms as shown in **Fig 13**, shows the step like behaviour. The different jump length is related with different diffusion pathways and their frequency. The number of such jumps will be related to the energy barrier for such diffusion pathways. In all the three compounds we observed that the least jump length is ~4-6 $Å^2$, which corresponds to Li migration from one tetrahedral site to another tetrahedral site (**Fig. 13**) directly i.e. along (100) as well as through octahedral site i.e. along (111) direction. The number of such jumps in all the three compounds is very large. This means that diffusion in all three compounds have significant contribution through Li jump along (100) and (111) directions. We have not found any interstitial position for Li occupancy. Hence in the crystalline phase the Li atoms diffuse only via tetrahedral jumps.

In **Fig. 14**, we have shown the Li trajectory of a few representative Li-ions in $Li_2X$(X=O,S and Se) compounds. This is useful to understand the minimum energy pathways of Li diffusion in these compounds. Our previous analysis (**Figs 12 and 13**) based on time dependence of mean squared displacement showed that only the Li- ions are diffusing while the X(O,S and Se) atoms form a rigid structure even at superionic temperature. Hence for easy visualization and explanation, we have shown (**Fig. 14)** only Li positions.

At ambient temperature lithium atoms occupy only tetrahedral sites (green spheres in **Fig 14**) in the crystal. However at higher temperature lithium may also jump from one tetrahedral to another tetrahedral site through the octahedral sites (yellow spheres in **Fig 14**). The time dependent positions of Li atoms are shown by colored dots in **Fig 14**. In case of $Li_2O$, Li atoms are hoping from one tetrahedral site to another tetrahedral site (green sphere to another green sphere) along (100) or its equivalent directions, i.e. they do not involve any intermediate octahedral position to hop from one tetrahedral site to another site. This is what we have inferred from our lattice dynamics analysis and nudged elastic band energy calculation that the easiest direction for Li migration would be along (100) direction.

Interestingly the story is very different for $Li_2S$ and $Li_2Se$. Here Li hops (**Fig. 14**) from one tetrahedral site to another tetrahedral site via octahedral sites as well as via direct jump along (100). But numbers of intermediate octahedral transient states are significant in these two compounds. Further, it is also to be noted that when one Li- hops from one tetrahedral site to another tetrahedral site, the Li at another site simultaneously jumps to another octahedral or tetrahedral site. So there is a cooperative hoping dynamics for $Li_2S$ and $Li_2Se$. Even in case of $Li_2O$ there is cooperative hoping but it seems that even for cooperative dynamics (100) is most favorable pathway.

The barrier energy which is obtained from nudged elastic band method (**Fig 8**) does not give the correct picture of barrier or activation energy for diffusion since NEB diffusion along (111) is prohibited due to large barrier for Li ion migration. Hence in order to correctly predict the diffusion barrier energy one may also need to see the correlated dynamics. Molecular dynamics simulation includes correlated as well as uncorrelated jumps, hence it truly represents the diffusion behaviour.

The isotropic diffusion coefficient is estimated using Einstein law of diffusion i.e. from the time dependence of mean square displacement as given below:

$$D=<u^2>/6\tau \qquad (5)$$

Where, $<u^2>$ is the change in the mean square displacement in time $\tau$.

The calculated diffusion coefficient as a function of temperature is shown in Fig 15. One could see that in $Li_2O$, the diffusion coefficient increases above 1200 K. However, in $Li_2S$ and $Li_2Se$ the diffusion starts at above 1000K and 900 K respectively. Further in order to correctly estimate the barrier energy from correlated as well uncorrelated dynamics for lithium diffusion in these compounds, we have fitted the temperature dependence of diffusion coefficients with Arhenius relation i.e.

$$D(T) = D_0 exp(-E_a/K_BT) \qquad (6)$$

Here $D_0$ is the constant factor, $K_B$ is the Boltzmann constant and T is temperature. One can linearize this equation by taking log of it, i.e.,

$$\ln(D(T)) = \ln(D_0) - E_a/K_B T \quad (7)$$

The conductivity measurement on Li$_2$S showed two different energy barriers of 0.70 eV and 1.52 eV in temperature range below and above 750 K respectively[23], while in Li$_2$O the values are 1.0 eV and 2.5eV in the temperature range below and above 1200 K respectively[66]. The high temperature value in the experimental analysis might be due to formations of defects and vacancy at high temperature in the superionic regime. There is no data available for Li$_2$Se. In Fig 15, we have fitted equation (7) and obtained the value of barrier energy for Li diffusion in all three compounds. The values are 0.98 eV, 0.79 eV and 0.87 eV for Li$_2$O, Li$_2$S and Li$_2$Se respectively.

## V. Conclusions

In this article we have reported extensive ab-initio lattice and molecular dynamics simulations aspect to investigate and understand the mechanism of Li diffusion in Li$_2$X anti-fluorite structures. The uncorrelated or single Li hopping from one tetrahedral to another tetrahedral site favors the <100> path, while <111> direction is least possible. However, correlated diffusion of Li is equally possible along <100> and <111> directions. Hence, for better understanding of pathways, the analysis of single ion dynamics through NEB method is not sufficient. However, one can include various correlated possibilities in the NEB method or use molecular dynamics simulation to evaluate the true barrier energy. The calculated barrier energy of Li$_2$S is the least among the three compounds; however, from the host stability point of view, Li$_2$O is a better candidate. By appropriate stoichiometry, one can trade off between these properties.

FIG. 1 (Color online) (a) Measured temperature dependence of inelastic neutron scattering spectra of $Li_2O$. For clarity the experimental data at various temperatures are shifted vertically. (b) The calculated neutron-cross-section weighted phonon density of states using lattice dynamics and compared with the measurement at T=300 K.

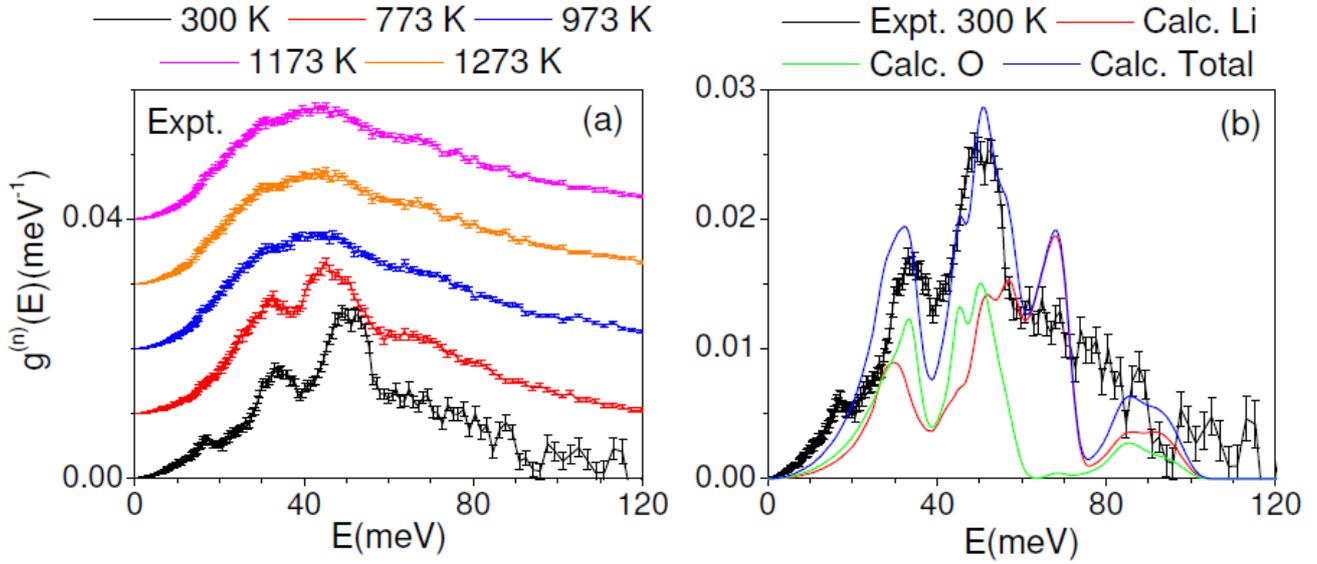

FIG. 2 (Color online) The calculated and experimental neutron weighted phonon density of states at 300 and 1200K. The phonon spectra have been calculated using ab-initio molecular dynamics simulations.

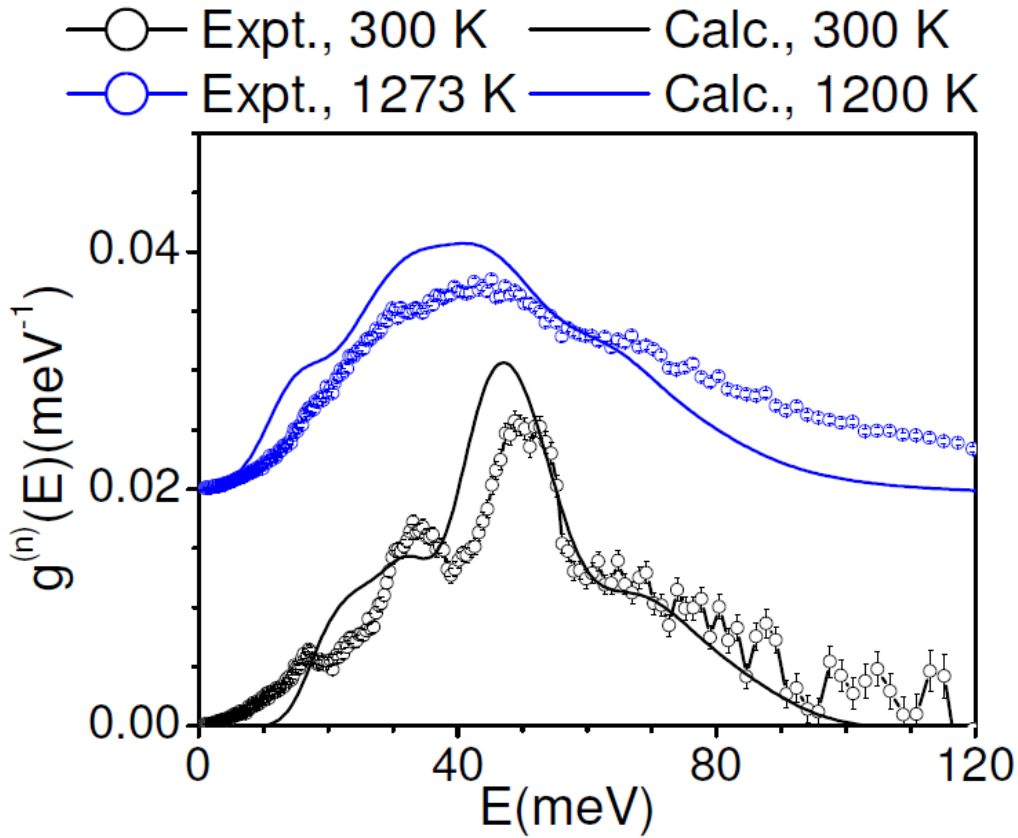

FIG. 3 (Color online) The calculated dispersion relation of $Li_2X$ (X=O, S and Se) along various high symmetry direction in Brillouin zone and compared with the available measurements. The solid lines and open circles correspond to the calculations and experimental data[62, 63].

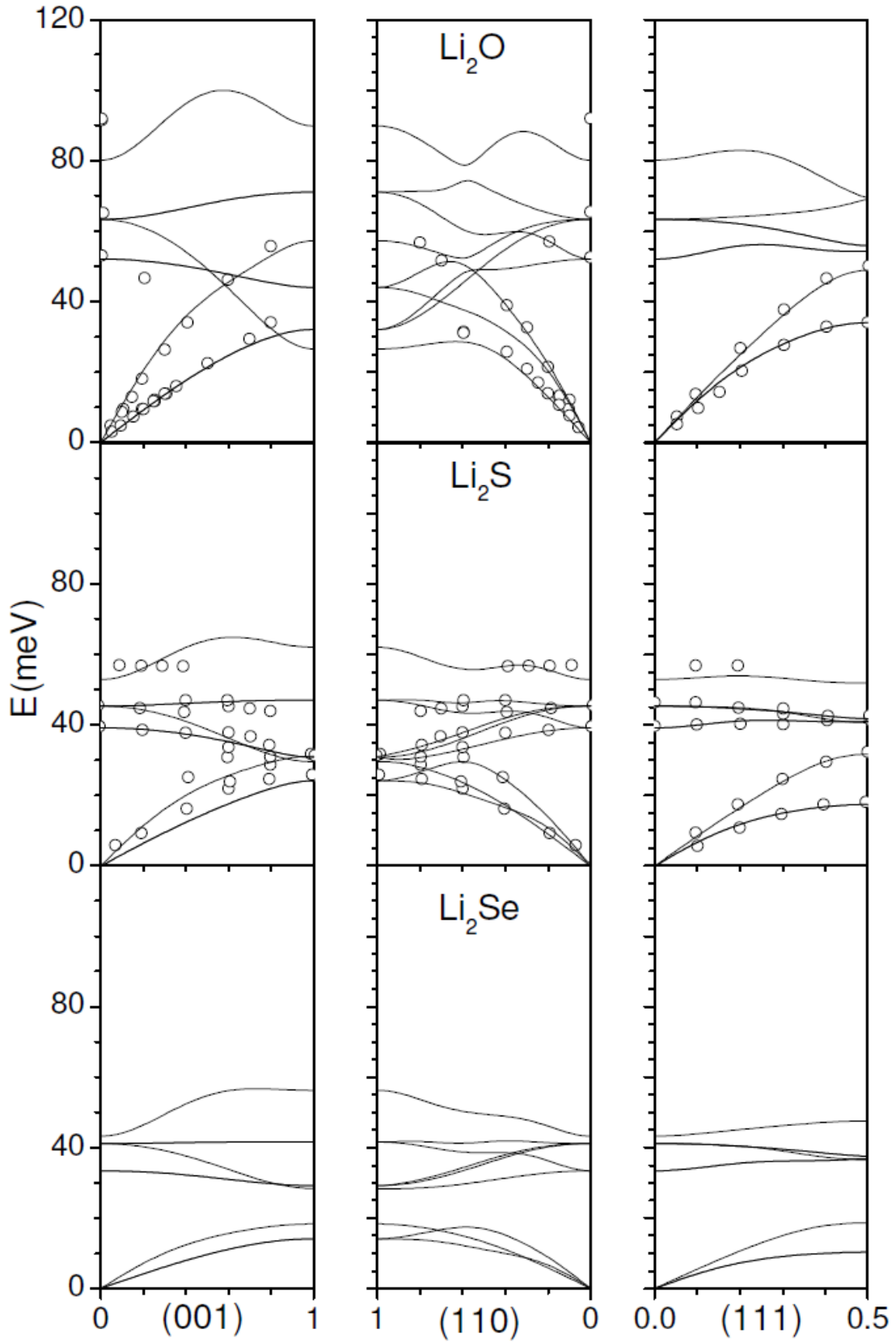

FIG. 4 (Color online) Cell parameter dependence of zone boundary phonon mode at (001) in $Li_2X$ (X=O, S and Se).

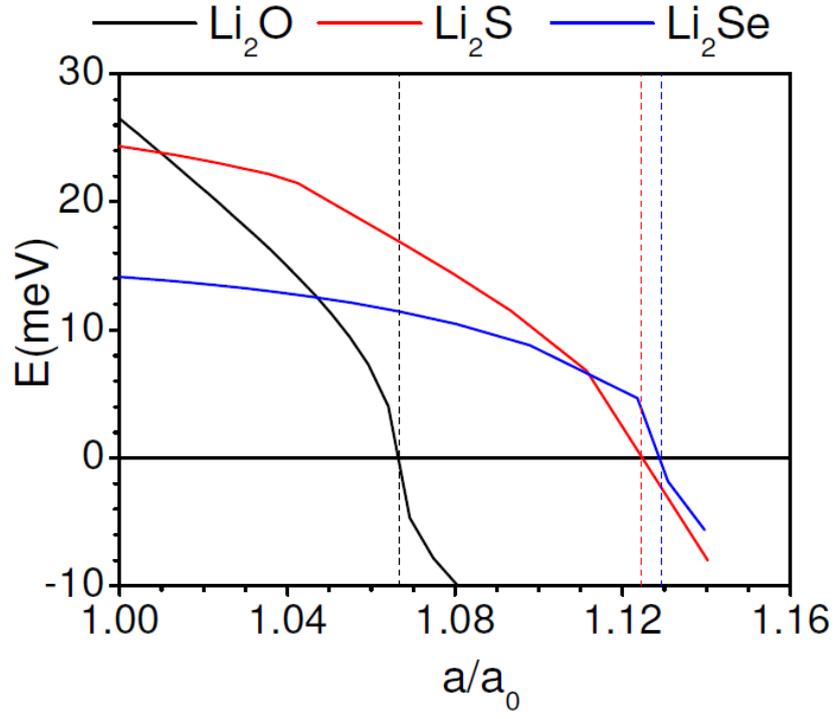

FIG. 5 (Color online) The lattice dynamics calculated total density of states and partial density of states of Li and X in $Li_2X$ (X=O, S, Se) compounds.

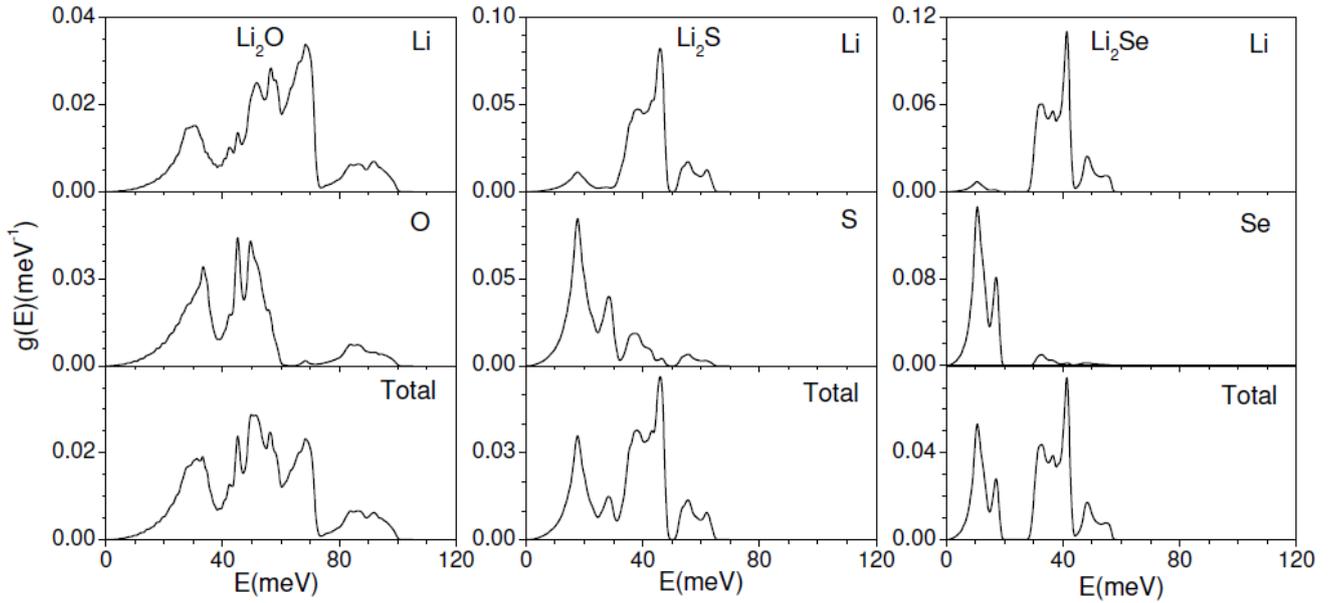

FIG. 6 (Color online) The calculated volume thermal expansion coefficient of $Li_2X$ (X=O,S and Se) compounds.

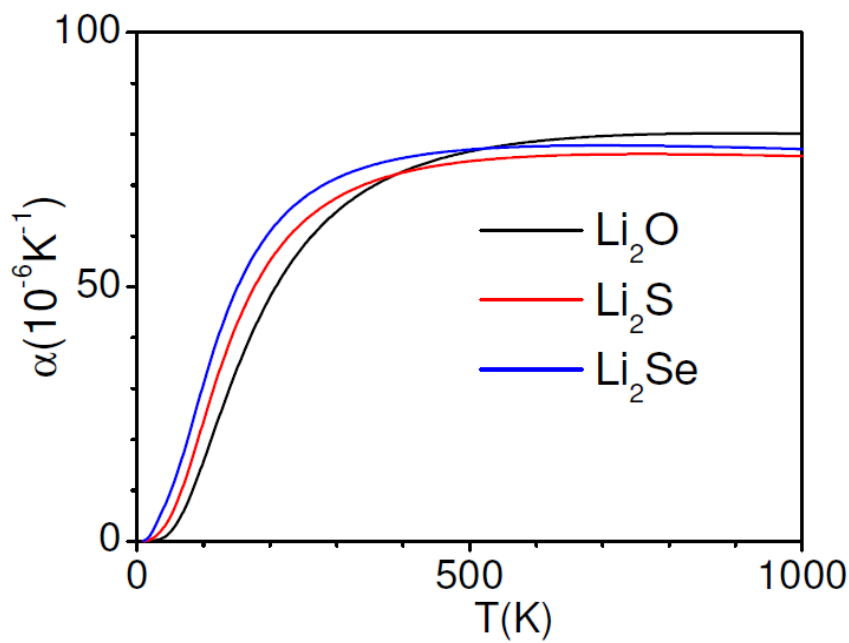

FIG. 7 (Color online) The calculated and experimental[64, 65] fractional change in lattice parameter with temperature of $Li_2X$ (X=O, S and Se).

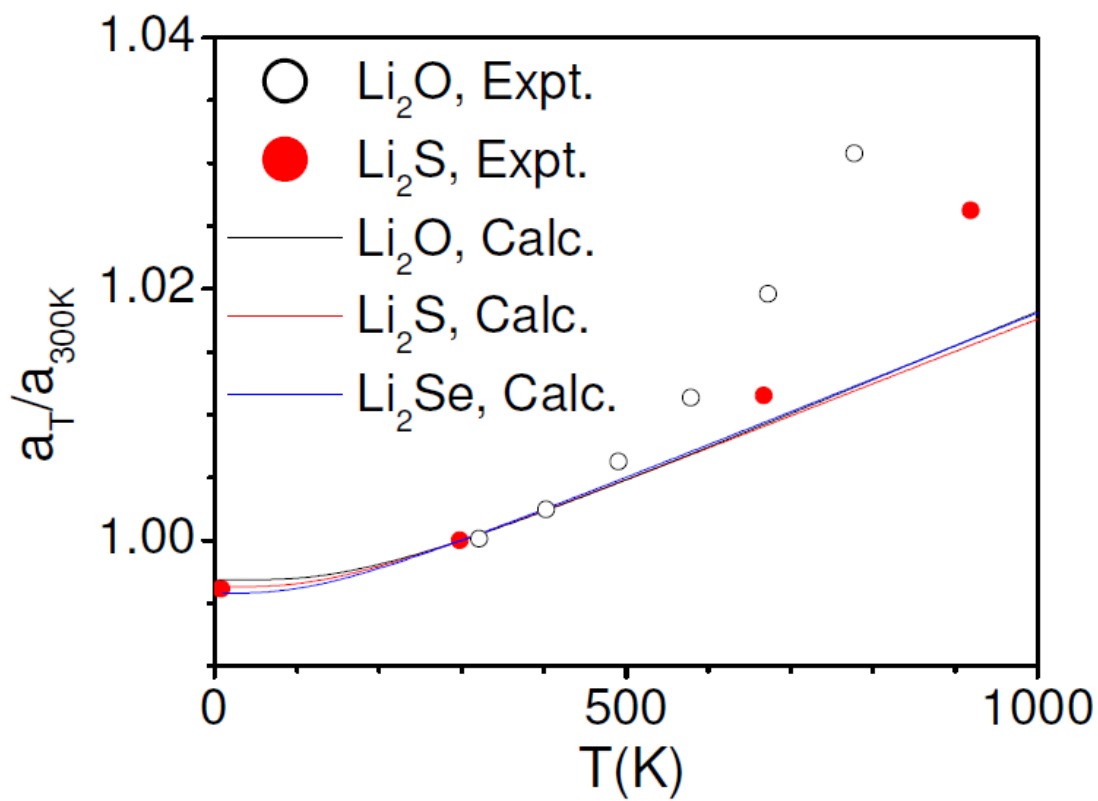

FIG. 8 (Color online) The calculated activation energy barrier for Li diffusion along various directions in the unit cell of $Li_2X$(X=O, S, Se). PATH-I (along [100]), PATH-II(along [110]), and PATH-III (along [111]) corresponds to correlated movement of 2, 2 and 4 Li atoms, respectively. PATH-III calculation are done in a 2×2×2 supercell. At ambient conditions lithium atoms occupy only tetrahedral sites (green spheres) in the crystal. The X (=O, S, Se) atoms are shown by red spheres.

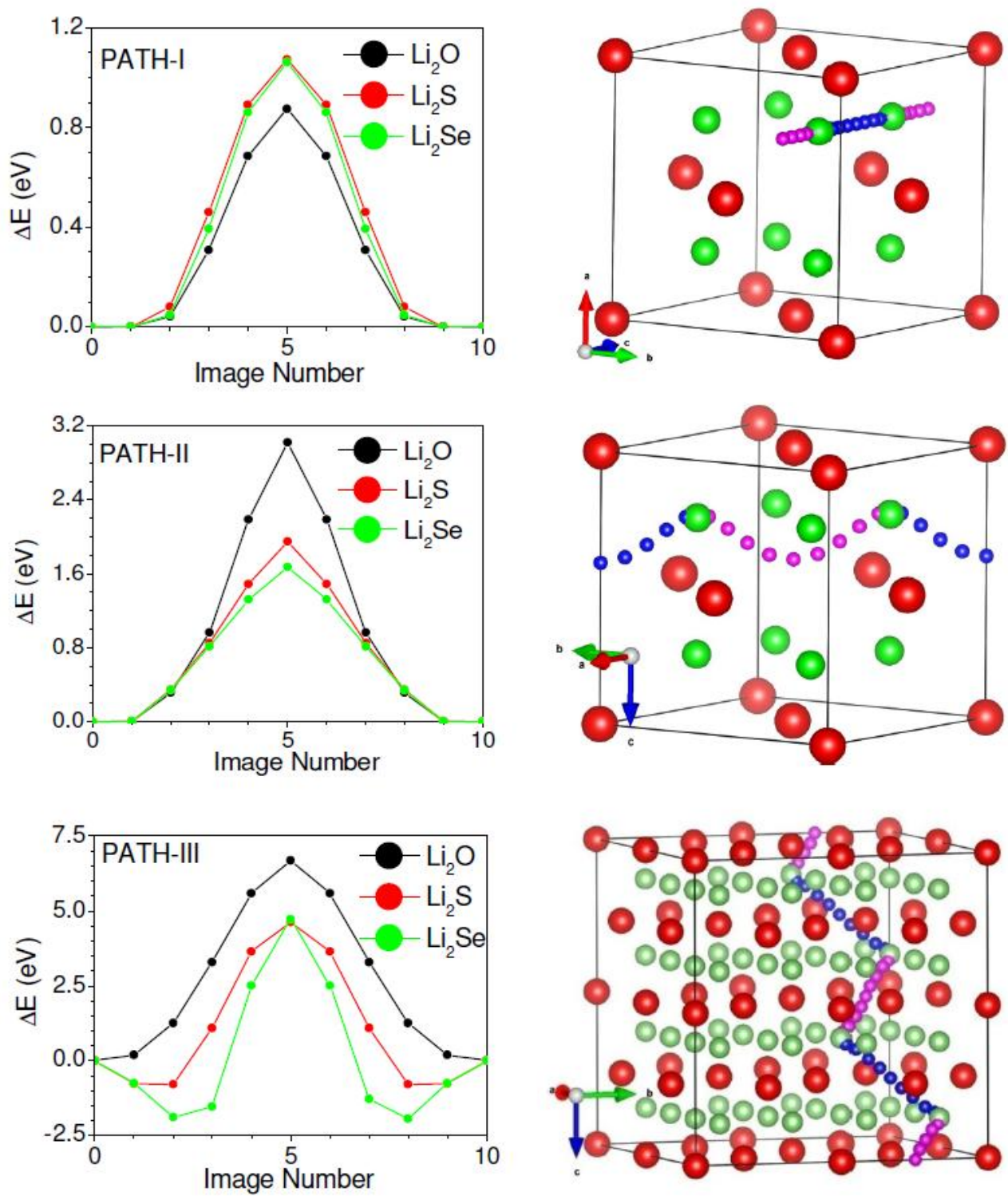

FIG. 9 (Color online) The calculated phonon spectrum of Li and X in $Li_2X$ (X=O, S and Se) compounds using molecular dynamics simulation at 300 K and above superionic transition temperature (T=1200 K).

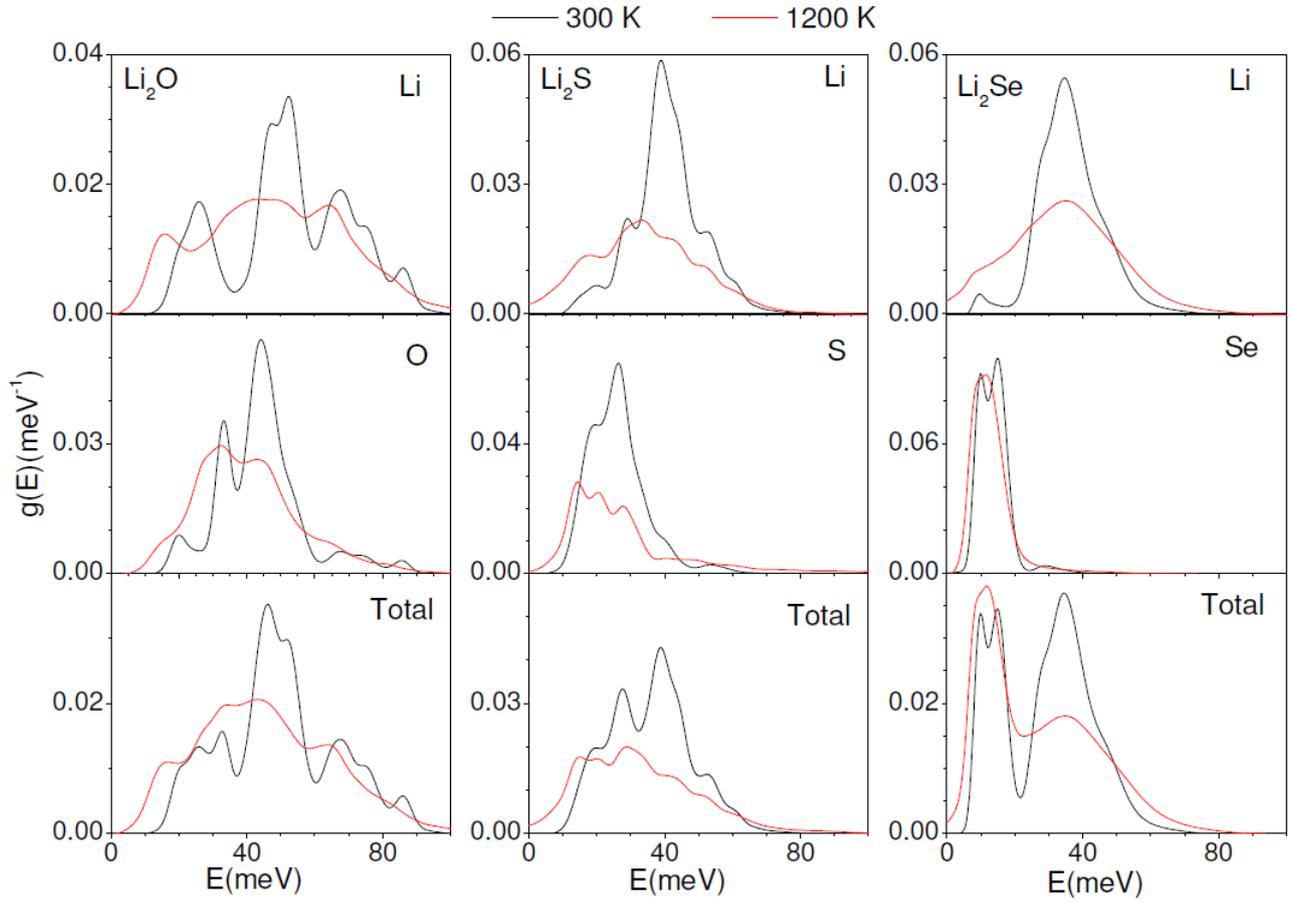

FIG. 10 (Color online) The calculated pair correlation function between various pairs of atom in $Li_2X$ (X=O, S and Se) at T=300 (solid line) and T=1200 K (dashed line).

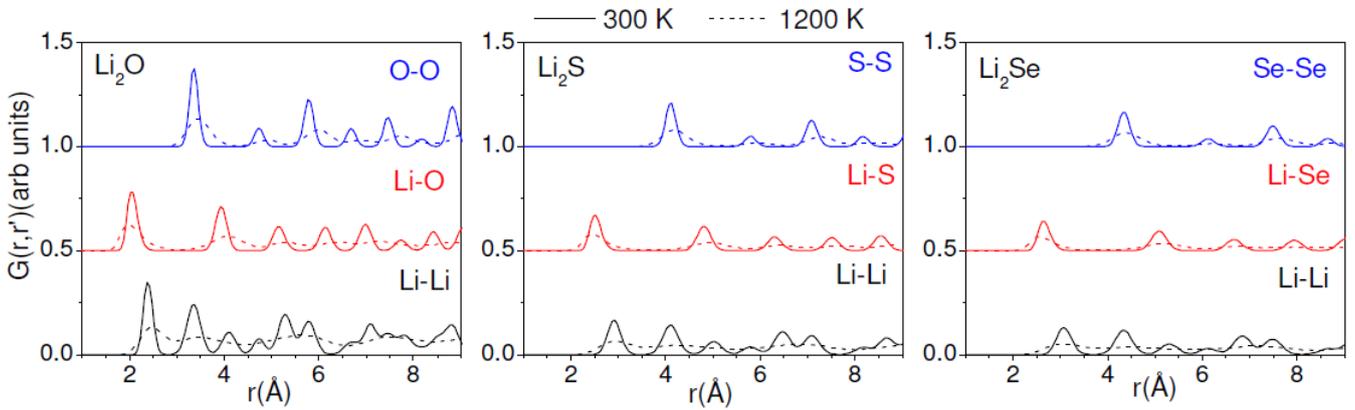

FIG. 11 (Color online) The calculated angle distribution between Li-X-Li junction in Li$_2$X(X=O,S and Se) at T=300 and T=1200 K averaged over 40 picosecond simulation time and all the atoms in supercell.

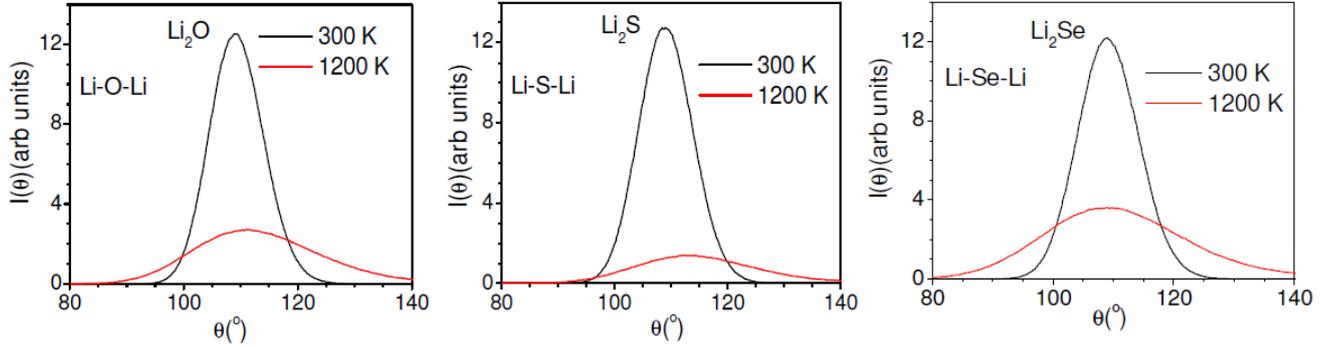

FIG. 12 (Color online) The calculated mean square displacement of Li and X atom in Li$_2$X (X=O, S and Se). The black and red solid lines corresponds to Li and X atoms.

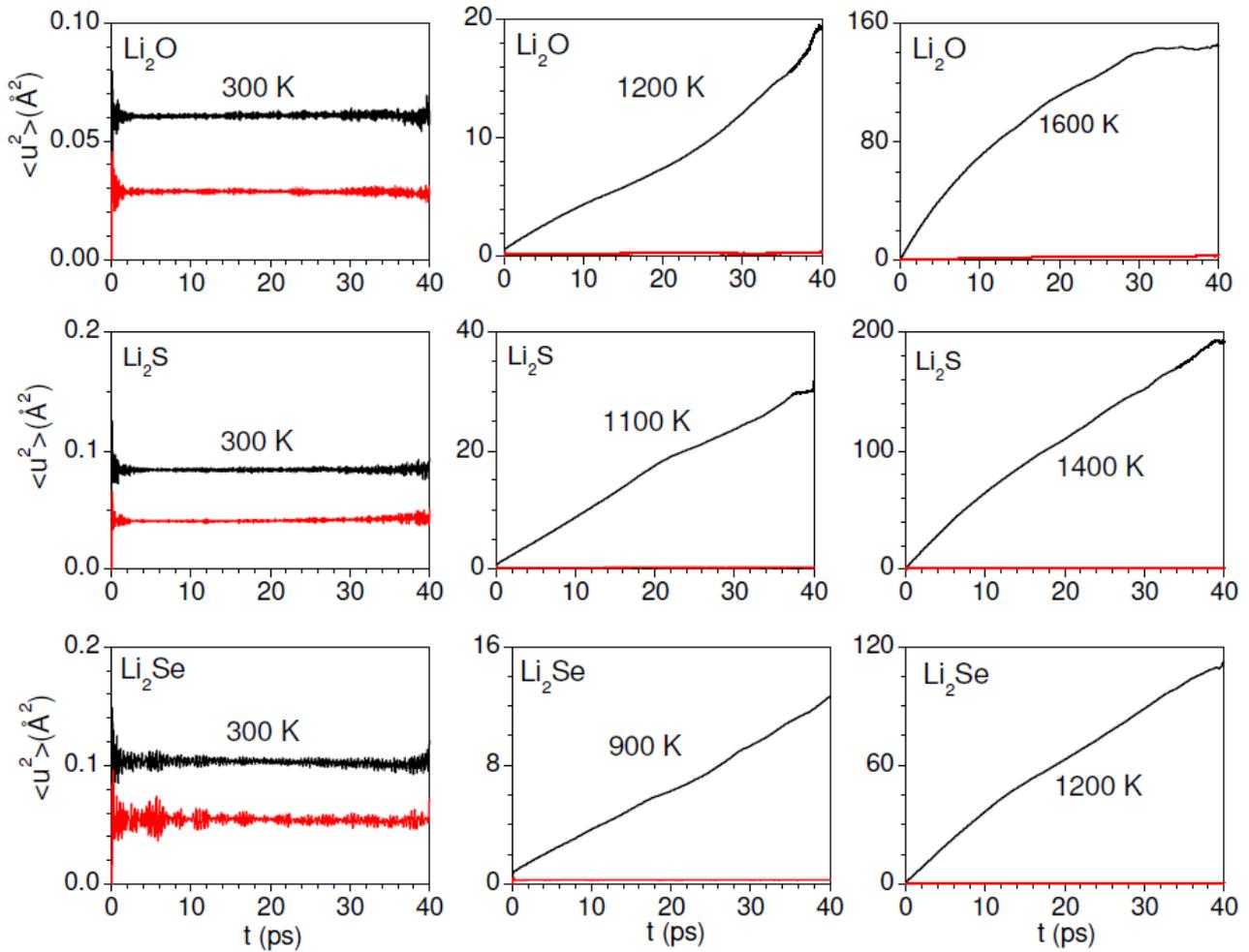

FIG. 13 (Color online) The calculated displacement of Li atoms in Li$_2$X (X=O,S and Se) at T=1200 K.

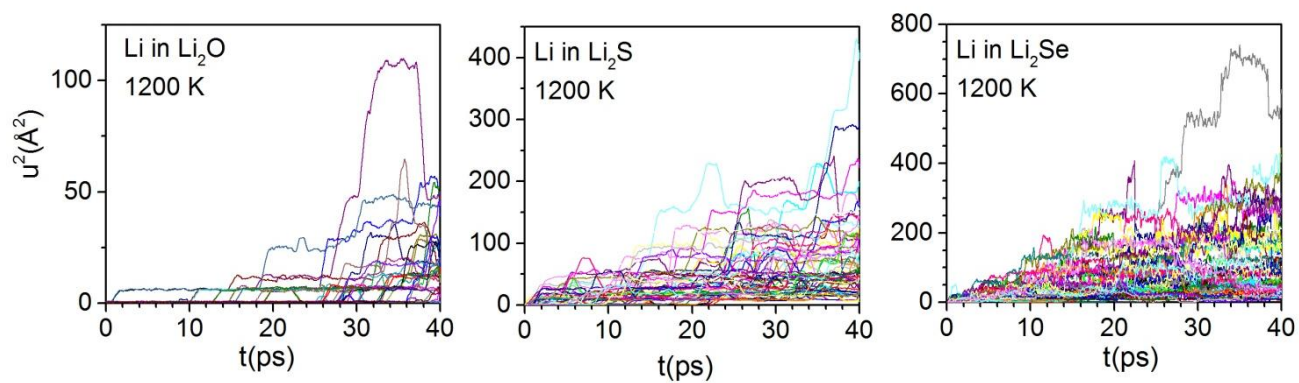

FIG.14 (Color online) The calculated trajectories of selected Li atoms at 1200 K in Li$_2$X (X=O, S and Se) in the a-b plane. At ambient conditions lithium atoms occupy only tetrahedral sites (green spheres) in the crystal. However at higher temperature lithium may also jump from one tetrahedral to another tetrahedral site through the octahedral sites (yellow spheres). X atoms are not shown. The time dependent positions of Li atoms are shown by colored dots. The calculated trajectories are shown on a 2×2×2 supercell.

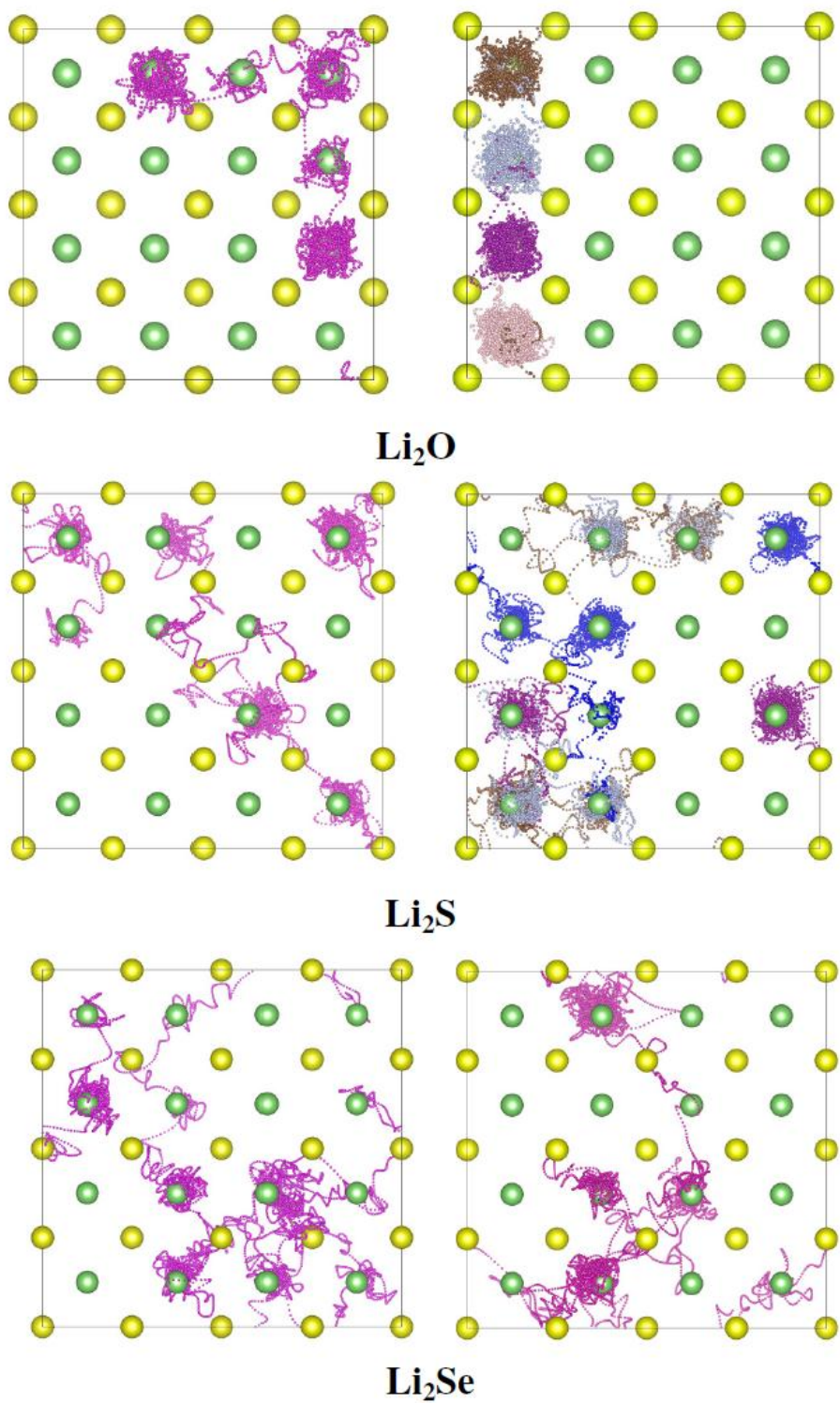

Fig. 15 (Color online) The calculated diffusion coefficients and activation energy barriers in Li$_2$X (X=O,S and Se).

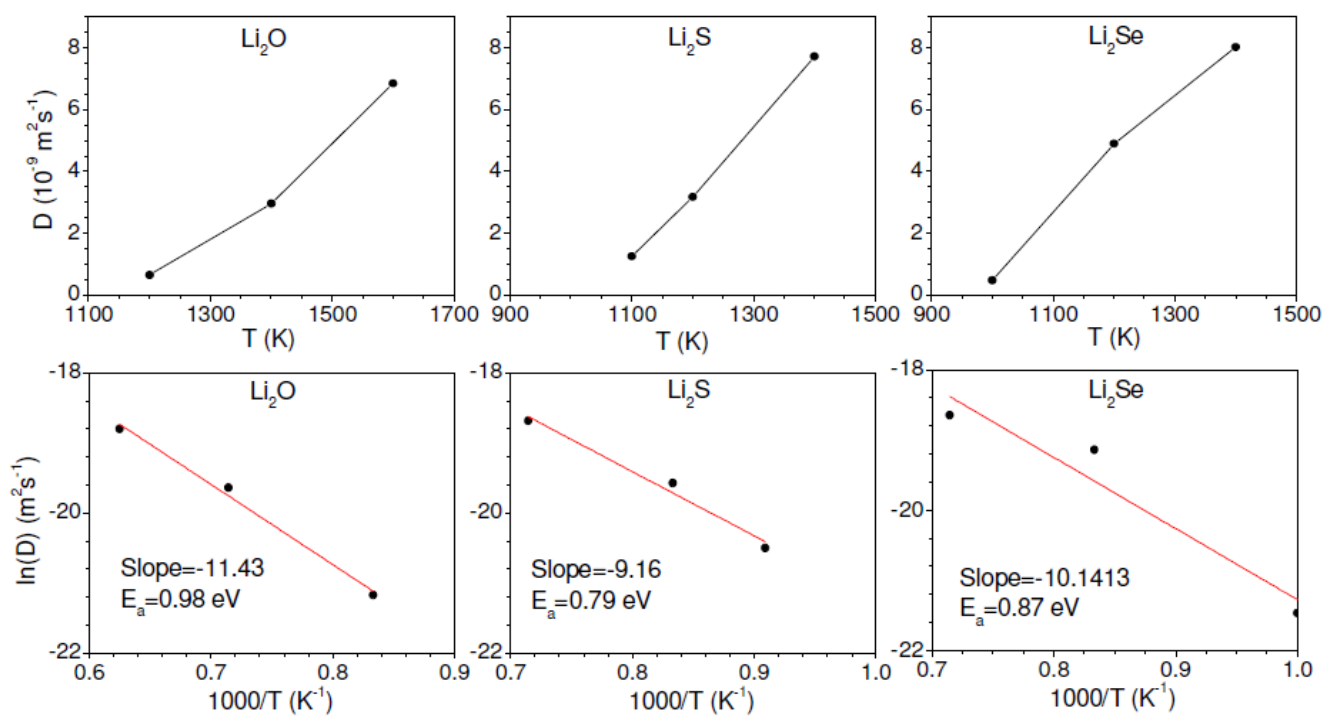